\documentclass[a4paper,12pt]{article}
\pdfoutput=1
\usepackage{feynmp-auto}
\usepackage{amsmath, amsfonts, amssymb}
\usepackage{graphicx}
\usepackage{enumerate}
\usepackage{hyperref}
\usepackage{latexsym}
\usepackage{dsfont}
\usepackage{hepnicenames}
\usepackage{enumerate}
\usepackage{soul}
\usepackage[normalem]{ulem}

\usepackage{units}

\usepackage{mathrsfs,graphicx,rotating,amsmath,amsfonts,mathtools,booktabs,amssymb,wasysym}
\usepackage{hyperref}\usepackage{slashed}
\usepackage[nosort]{cite}
\usepackage[table,xcdraw,dvipsnames]{xcolor}
\usepackage{bm}
\usepackage{graphicx}
\usepackage{multirow,multicol}
\hypersetup{colorlinks,bookmarksopen,bookmarksnumbered,
linkcolor=blus,pdfstartview=FitH,urlcolor=rossos,citecolor=verde}
\allowdisplaybreaks

\renewcommand{\[}{\left[}

\def\Lag{\mathscr{L}}

\newcommand{\mio}[1]{}

 \newcommand{\med}[1]{\langle #1\rangle}

\def\bpm{\begin{pmatrix}}
\def\epm{\end{pmatrix}}

\usepackage{mathrsfs}

 \newcommand{\fig}[1]{~\ref{fig:#1}}
\newcommand{\sfrac}[2]{#1/#2}

\allowdisplaybreaks
\usepackage{multicol}
\usepackage{color}
\definecolor{rosso}{cmyk}{0,1,1,0.4}
\definecolor{rossos}{cmyk}{0,1,1,0.55}
\definecolor{rossoc}{cmyk}{0,1,1,0.2}
\definecolor{blu}{cmyk}{1,1,0,0.3}
\definecolor{blus}{cmyk}{1,1,0,0.6}
\definecolor{bluc}{cmyk}{1,1,0,0.1}
\definecolor{verde}{cmyk}{0.92,0,0.59,0.25}
\definecolor{verdec}{cmyk}{0.92,0,0.59,0.15}
\definecolor{verdes}{cmyk}{0.92,0,0.59,0.4}

\newcommand{\eq}[1]{~{\rm (\ref{eq:#1})}}

\newcommand{\MeV}{\,{\rm MeV}}
\newcommand{\GeV}{\,{\rm GeV}}
\newcommand{\TeV}{\,{\rm TeV}}
\newcommand{\cm}{\,{\rm cm}}

\def\circa#1{\,\raise.3ex\hbox{$#1$\kern-.75em\lower1ex\hbox{$\sim$}}\,}

\newcommand{\beq}{\begin{equation}}
\newcommand{\eeq}{\end{equation}}

\newcommand{\bea}{\begin{eqnarray}}
\newcommand{\eea}{\end{eqnarray}}
\newcommand{\be}{\begin{equation}}
\newcommand{\ee}{\end{equation}}
\font\tenrsfs=rsfs10 at 12pt
\font\sevenrsfs=rsfs7
\font\fiversfs=rsfs5
\newfam\rsfsfam
\textfont\rsfsfam=\tenrsfs
\scriptfont\rsfsfam=\sevenrsfs
\scriptscriptfont\rsfsfam=\fiversfs

\newsavebox\MBox

\newcommand{\eV}{\,{\rm eV}}
\newcommand{\SU}{\,{\rm SU}}

\def\circa#1{\,\raise.3ex\hbox{$#1$\kern-.75em\lower1ex\hbox{$\sim$}}\,}
\makeatletter

\font\ital=cmu10

\def\hhref#1{\href{http://arxiv.org/abs/#1}{arXiv:#1}}
\usepackage{xstring}
\newcommand{\hhrefq}[1]{\IfSubStr{#1}{:}{\href{http://inspirehep.net/search?ln=en&ln=en&p=#1&of=hb&action_search=Search&sf=&so=d&rm=&rg=25&sc=0}{InSpire:#1}}{\hhref{#1}}}

\def\art{\@ifnextchar[{\eart}{\oart}}
\def\eart[#1]#2#3#4#5#6{{\rm #2}, {\em #3 \bf #4} {\rm (#6) #5} ({\em #1})}
\def\article{\@ifnextchar[{\earticle}{\oarticle}}
\def\oarticle#1#2#3#4#5#6{{\rm #1}, {\ital ``#6''}, {\rm #2 #3 (#5) #4}}
\def\earticle[#1]#2#3#4#5#6#7{{\rm #2}, {\ital ``#7''}, {\rm #3 #4 (#6) #5}  [\hhrefq{#1}]}
\def\hepart[#1]#2{{\rm #2, \sl#1}}
\def\heparticle[#1]#2#3{#2, {\ital ``#3''} [\hhrefq{#1}]}
\newcommand{\doi}[1]{\href{http://dx.doi.org/#1}{[link]}}

\renewenvironment{thebibliography}[1]
     {\begin{multicols}{2}[\section*{\refname}]%
      \@mkboth{\MakeUppercase\refname}{\MakeUppercase\refname}%
      \list{\@biblabel{\@arabic\c@enumiv}}%
           {\settowidth\labelwidth{\@biblabel{#1}}%
            \leftmargin\labelwidth
            \advance\leftmargin\labelsep
            \@openbib@code
            \usecounter{enumiv}%
            \let\p@enumiv\@empty
            \renewcommand\theenumiv{\@arabic\c@enumiv}}%
      \sloppy
      \clubpenalty4000
      \@clubpenalty \clubpenalty
      \widowpenalty4000%
      \sfcode`\.\@m}
     {\def\@noitemerr
       {\@latex@warning{Empty `thebibliography' environment}}%
      \endlist\end{multicols}}

\newcounter{alphaequation}[equation]
\def\thealphaequation{\theequation\hbox to
0.6em{\hfil\alph{alphaequation}\hfil}}
\def\eqnsystem#1{
\def\@eqnnum{{\rm (\thealphaequation)}}
\def\@@eqncr{\let\@tempa\relax \ifcase\@eqcnt \def\@tempa{& & &} \or
  \def\@tempa{& &}\or \def\@tempa{&}\fi\@tempa
  \if@eqnsw\@eqnnum\refstepcounter{alphaequation}\fi
\global\@eqnswtrue\global\@eqcnt=0\cr}
\refstepcounter{equation} \let\@currentlabel\theequation \def\@tempb{#1}
\ifx\@tempb\empty\else\label{#1}\fi
\refstepcounter{alphaequation}
\let\@currentlabel\thealphaequation
\global\@eqnswtrue\global\@eqcnt=0 \tabskip\@centering\let\\=\@eqncr
$$\halign to \displaywidth\bgroup \@eqnsel\hskip\@centering
$\displaystyle\tabskip\z@{##}$&\global\@eqcnt\@ne
\hskip2\arraycolsep\hfil${##}$\hfil& \global\@eqcnt\tw@\hskip2\arraycolsep
$\displaystyle\tabskip\z@{##}$\hfil
\tabskip\@centering&\llap{##}\tabskip\z@\cr}
\def\endeqnsystem{\@@eqncr\egroup$$\global\@ignoretrue} \makeatother

\definecolor{Gray}{gray}{0.95}

\def\bal#1\eal{\begin{align}#1\end{align}}

\begin{document}

{ULB-TH/18-06 \hfill CERN-TH-2018-110\hfill IFUP-TH/2018}

\vspace{1.5cm}

\begin{center}
{\Large\LARGE \bf \color{rossos}
Super-cool Dark Matter}\\[1cm]
{\bf Thomas Hambye$^a$, Alessandro Strumia$^{b,c,d}$, Daniele Teresi$^a$}\\[7mm]

{\it $^a$ Service de Physique Th\'eorique, Universit\'e Libre de Bruxelles, 
Brussels, Belgium}\\[1mm]
{\it $^b$ CERN, Theory Division, Geneva, Switzerland}\\[1mm]
{\it $^c$ Dipartimento di Fisica dell'Universit{\`a} di Pisa}\\[1mm]
{\it $^d$ INFN, Sezione di Pisa, Italy}\\[1mm]

\vspace{0.5cm}

{\large\bf\color{blus} Abstract}
\begin{quote}\large
In dimension-less theories of dynamical generation of the weak scale,
the Universe can undergo a period of low-scale inflation during which all particles are massless
and undergo super-cooling.
This leads to a new mechanism of generation of the cosmological Dark Matter  relic density:
super-cooling can easily suppress the amount of Dark Matter down to the desired level.
This is achieved for TeV-scale Dark Matter, if super-cooling ends when quark condensates form at the QCD phase transition.
Along this scenario, the baryon asymmetry can be generated either at the phase transition or
through leptogenesis.
We show that the above mechanism  takes place in old and new dimension-less models.
\end{quote}

\thispagestyle{empty}
\bigskip

\end{center}

\setcounter{footnote}{0}

\newpage

\section{Introduction}
Dark Matter (DM) could be particles with mass $M_{\rm DM} \gg {\rm keV}$.
The cosmological DM abundance $\Omega_{\rm DM} h^2 \approx 0.110$
is reproduced  if their number density
in units of the entropy density $s$ is small:
\beq \label{eq:0.40eV}
\frac{n_{\rm DM}}{s} = \frac{0.40\eV}{M_{\rm DM}} \frac{\Omega_{\rm DM} h^2 }{0.110}. \eeq
When DM becomes non-relativistic,
thermal freeze-out at $T \sim T_{\rm dec}\approx M_{\rm DM}/25$
 leaves the  DM abundance 
$ \sfrac{n_{\rm DM}}{s} \sim 1/(M_{\rm DM} M_{\rm Pl} \sigma_{\rm ann})$,
where $\sigma_{\rm ann}$ is the DM annihilation cross section.
As well known, $M_{\rm DM} \sim \TeV$ and $\sigma_{\rm ann} \sim 1/M_{\rm DM}^2$  reproduces the desired DM abundance.
Many alternative cosmological DM production mechanisms are possible,
sometimes at the price of increasing model-building complexity.

We here discuss a new mechanism that can generate the desired cosmological DM abundance.
The new mechanism is characteristic of models where a  scale
(we will consider the weak scale) is dynamically generated from a quantum field theory
that only has dimension-less couplings.
We assume that, in this context, all particles (in particular dark matter and the Higgs boson) remain massless until a vacuum expectation value
or condensate develops.  In the case of scalars, this conjecture
is at odds with the usual view that attributes physical meaning
to power-divergent quantum corrections, leading to the expectation that the Higgs boson should have
been accompanied by new physics able of keeping its mass  naturally much smaller than the Planck mass,
the presumed cut-off of quantum field theories.
The observation of the Higgs boson not accompanied by any new physics promoted renewed interest in dimension-less dynamics
as the possible origin of the weak scale, see e.g.~\cite{Bardeen:1995kv,0710.2840,1303.7244,1603.03603,1804.06376},
and in attempts of building weak-scale extensions of the Standard Model valid up to infinite energy,
such that no cut-off is needed~\cite{1403.4226,1412.2769,1507.06848,1705.03896}.
More generically, super-cooling takes place if, for whatever reason, the
mass scales in the potential are much smaller than the scale  generated trough dynamical transmutation.
If small scalar masses are unnatural, 
super-cooling is a cosmological signature of such unnaturalness.

The mechanism relies on the fact that in dimension-less models, along its thermal history, the Universe remains trapped for a while in a phase of thermal inflation
during which all particles are massless, so that DM undergoes super-cooling rather than freeze-out.
The formation of QCD condensates at the QCD phase transition ends this phase, leading to the electro-weak phase transition, i.e.~particle mass generation.
From this point particles lighter than the reheating temperature can easily thermalize, but DM will not necessarily thermalize, leading to the necessary suppression of the DM relic density,
provided that its mass is at the TeV scale.
In the context of freeze-out, the same coincidence is advertised as `WIMP miracle'.

In section~\ref{general} we present the mechanism.
In the next sections we consider specific models, pointing out that no ad hoc model building is needed. 
Indeed, in section~\ref{specific} we consider the model proposed in~\cite{Hambye:2013sna} (for the non scale invariant version of this model see \cite{Hambye:2008bq,Hambye:2009fg,Arina:2009uq}),
that extends the SM by adding a scalar doublet under a new $\SU(2)$ gauge group (its vectors are automatically stable DM candidates).
We find that this model can reproduce the observed DM either through freeze-out (at larger values of the gauge coupling~\cite{Hambye:2013sna})
or through super-cooling (at smaller values of the gauge coupling).
Super-cooling erases the baryon asymmetry: we will discuss how it can be regenerated at the weak scale, possibly through leptogenesis.
Motivated by leptogenesis and neutrino masses, in section~\ref{B-L} we propose a similar model with U(1)$_{B-L}$ gauge group and
two extra scalars. Conclusions are given in section~\ref{concl}.

While we focus on simple models based on weakly-coupled elementary particles,
super-cool DM can also arise in more generic contexts, such as
strongly coupled models with walking dynamics~\cite{Bando:1987br}, 
possibly described through broken conformal symmetries and/or through branes in warped extra dimensions~\cite{1104.4791}.
In such a case super-cooling can be described geometrically~\cite{hep-th/0107141}.
Furthermore, super-cool DM could arise in extensions of the Standard Model that cut quadratically divergent 
corrections to the Higgs mass and generate the weak scale: for example in supersymmetric models
where all particles are massless in the supersymmetric limit, and where supersymmetry gets dynamically broken
by some expectation value.  However, the non-observation of new physics at the Large Hadron Collider casts doubts on such models. In more general terms, we expect the mechanism discussed here to be active whenever DM predominantly acquires its mass in a phase transition occurring after a significant period of super-cooling.

\section{General mechanism}\label{general}
We consider extensions of the SM that provide a DM candidate and where  all particles get mass from
the vacuum expectation value of a scalar $s$, sometimes called `dilaton'.
In the standard freeze-out scenario, DM with mass $M_{\rm DM}$ would decouple at a temperature
$T_{\rm dec}$, equal to $T_{\rm dec} \approx M_{\rm DM}/25$ if freeze-out
reproduces the cosmological DM density.

\subsection{Super-cooling}
In dimension-less models, due to the absence of quadratic terms and vacuum stability (i.e.\ positive scalar quartics),
thermal effects 
select, as the Universe cools down, the false vacuum where all scalars
(in particular the dilaton $s$ and the Higgs $h$) have vanishing vacuum expectation values. Around this vacuum,
all particles are massless, including DM.
The energy density of the Universe receives two contributions:
from radiation, $\rho_{\rm rad}(T)= \sfrac{g_* \pi^2 T^4}{30}$,
and from the vacuum energy of the false vacuum $V_\Lambda >0$.
We assume that the true vacuum has a nearly zero vacuum energy, as
demanded by the observed small cosmological constant.

While the Universe cools,
the vacuum energy starts dominating over radiation at some temperature $T = T_{\rm infl}$
starting a phase of thermal inflation with Hubble constant $H$, such that
$V_\Lambda$ determines $ T_{\rm infl}$ and $H$ as
\begin{equation}\label{eq:TinflH}
  \frac{g_* \pi^2 T^4_{\rm infl}}{30} =  V_\Lambda = \frac{3H^2M_{\rm Pl}^2}{8\pi}
\end{equation}
where $g_*$ is the number of relativistic degrees of freedom just before inflation starts.
We denote as $a_{\rm infl}$ the scale factor of the Universe at this stage
and write ${\cal O}(1)$ factors in Boltzmann approximation (correct within $\pm10\%$)
given that DM might be bosonic or fermionic.

During this phase DM is massless and thereby remains coupled, rather than
undergoing thermal freeze-out at $T \sim T_{\rm dec}$ as e.g.\ in~\cite{1507.08660}.
All particles, including DM, undergo super-cooling:
the scale factor of the Universe grows as $a = a_{\rm infl} e^{Ht}$, and the temperature
drops as $T = T_{\rm infl} a_{\rm infl}/a$.

Super-cooling ends at some temperature $T_{\rm end}$ with a phase transition towards 
the true vacuum at $\med{s},\med{h}\neq 0$ ---
at the end of this section we will discuss how this happens in the models of interest.
During thermal inflation, the scale factor of the Universe inflates  by a factor
$e^N = \sfrac{T_{\rm infl}}{T_{\rm end}} $.

\subsection{Reheating}
After the first-order phase transition to the true vacuum, the  various particles, including DM, become massive, and
the energy density $V_\Lambda$ stored in the scalars
is transferred to particles, reheating the Universe.
If the energy transfer  rate  $\Gamma$ is much faster than the Hubble rate $H$,
the reheating temperature is $g_{\rm RH}^{1/4}
T_{\rm RH}= g^{1/4}_* T_{\rm infl}$,
where $g_{\rm RH}$ is the number of reheated degrees of freedom.
Otherwise the scalars, before decaying, undergo a period of oscillations and the reheating temperature is lower.
During this period, the scalars dilute as matter.
When scalars finally decay, their remaining energy density $\rho_{\rm sca}$
becomes radiation with reheating temperature 
\beq T_{\rm RH}\approx T_{\rm infl} \min(1,\Gamma/H)^{1/2}.\eeq
The final DM abundance $Y_{\rm DM}= n_{\rm DM}/s$
(where $s = 2\pi^2 g_* T^3/45$ is the entropy density)
receives two contributions: 
\beq Y_{\rm DM} \approx  Y_{\rm DM}|_{\rm super-cool} +Y_{\rm DM}|_{\rm sub-thermal}.\label{twopopulations}\eeq
The super-cool contribution to the DM abundance is what remains
of the original  population of formerly massless DM in thermal equilibrium,
suppressed by the the dilution due to thermal inflation
\beq \label{eq:Ysc}
Y_{\rm DM}|_{\rm super-cool} =  Y^{\rm eq}_{\rm DM}   \frac{T_{\rm RH}}{T_{\rm infl}}
 \left(\frac{T_{\rm end}}{T_{\rm infl}}\right)^3,\qquad
 Y^{\rm eq}_{\rm DM}  =\frac{45g_{\rm DM}}{2\pi^4 g_*}  
 . \eeq
The factor  $T_{\rm RH}/T_{\rm infl}$ arises taking into account that
the energy stored in the oscillating inflaton dilutes as matter (rather than as radiation)
between the end of inflation and reheating, when it is finally converted to radiation.


The second contribution is the population of DM particles which can be produced from the thermal bath, through scattering effects, after reheating.\footnote{Production
during pre-heating was discussed in~\cite{1104.4791}.
Production during bubble collisions was discussed in~\cite{1211.5615}.}
If in this way DM has the time to thermalize again, i.e.\ if 
$T_{\rm RH} \gtrsim T_{\rm dec}$, the supercooled population is erased and this second population  reaches thermal equilibrium. 
In this case DM undergoes a usual freeze-out, leading to a relic density independent of whether there was previously a supercooling period.
If instead $T_{\rm RH} \lesssim T_{\rm dec}$ the super-cool population remains basically unchanged, and the second population 
is produced with a sub-thermal abundance.
This population is determined by the usual Boltzmann equation for
$Y_{\rm DM} = n_{\rm DM}/s$
in a radiation-dominated Universe, the same equation as the one that controls DM freeze-out.
In non-relativistic approximation
\beq  \frac{dY_{\rm DM}}{dz} = \frac{\lambda}{z^2} (Y^2 _{\rm DM}- Y_{\rm DM}^{\rm 2eq}),\qquad\hbox{where}\qquad
z = \frac{M_{\rm DM}}{T}.
\label{eq:Boltzf}\eeq
where
$ \lambda =
{M_{\rm Pl} M_{\rm DM} \langle\sigma_{\rm ann} v_{\rm rel}\rangle}\sqrt{\sfrac{\pi g_{\rm SM}}{45}}$
if DM annihilations have a cross section $\sigma_{\rm ann}$ dominated by $s$-wave scattering.
For $\lambda \gg 1$ freeze-out occurs at $T_{\rm dec} \approx M_{\rm DM}/ \ln\lambda$; otherwise DM never thermalizes again after reheating and the term proportional to $Y_{\rm DM}^{\rm 2eq}$ can be neglected.
Integrating eq.\eq{Boltzf} the regenerated DM population (starting from $T = T_{\rm RH}$) is
\beq \label{eq:YDMthermal}
Y_{\rm DM}|_{\rm sub-thermal} = \lambda \int_{z_{\rm RH}}^\infty \frac{dz}{z^2} Y_{\rm eq}^2 = \lambda  \, \frac{2025g_{\rm DM}^2}{128\pi^7 g_{\rm SM}^2}
e^{-2z_{\rm RH}} (1+2 z_{\rm RH}).\eeq

Summarising, the super-cool contribution dominates provided that $T_{\rm RH}\ll T_{\rm dec}$.
If DM interaction are small enough that it does not undergo kinetic recoupling, 
DM remains colder than in the freeze-out scenario: this has little observational implications~\cite{Gunn:1978gr};
the fact that super-cool DM forms smaller structures could give enhanced tidal fluctuations possibly observable
along the lines of~\cite{1710.06443}.
The super-cool population of eq.\eq{Ysc} does not depend on DM properties
(provided that initially, before supercooling, it was in thermal equilibrium), but only on the
amount of super-cooling.
As we will see below, eq.\eq{Ysc} matches the desired DM abundance provided that $T_{\rm end}$
is a few orders of magnitude below $T_{\rm RH}$. If instead  $T_{\rm RH} \lesssim T_{\rm dec}$, an additional non-thermal DM population is generated after reheating which can also lead to the desired relic density.
%

\subsection{The phase transition to the massive vacuum}
Super-cooling ends anyway
at the nucleation temperature $T_{\rm nuc}$ when the rate
of thermal vacuum decay becomes faster than the Hubble rate.
So far nothing connects the DM mass to the weak scale.
However, if this nucleation temperature lies below the QCD phase transition, 
super-cooling  in fact ends sooner, at this transition~\cite{Witten:1980ez,Iso:2017uuu}.
The temperature at which super-cooling ends, $T_{end}$ is  approximated by
\begin{equation}\label{eq:Tend}
T_{\rm end} = \max \bigg[T_{\rm nuc}, \min\bigg(T_{\rm cr}^{\rm QCD}, T_{\rm end}^{\rm QCD} \bigg) \bigg] \,.
\end{equation}
Indeed, quark condensates form at $T_{\rm cr}^{\rm QCD}\sim  \Lambda_{\rm QCD}$.
In view of its Yukawa couplings to quarks (in particular to the top)
the Higgs acquires a vacuum expectation value  $\med{h}_{\rm QCD} \sim \Lambda_{\rm QCD}$,
which induces a squared mass term $M_s^2$ for the $s$ scalar.
If $M_s^2$ is negative and bigger in modulus than the thermal $s$ mass
(which dominates the potential around the origin in dimension-less theories)
$s$ immediately starts rolling down, ending super-cooling at $T_{\rm cr}^{\rm QCD}$.
Otherwise, $s$  starts rolling at a lower temperature 
$T_{\rm end}^{\rm QCD}$, as soon as its thermal mass becomes smaller than $|M_s|$.
When super-cooling is stopped at $T_{\rm end}\sim \Lambda_{\rm QCD}$ the
amount of super-cooling that reproduces the observed DM abundance is obtained for a DM mass fixed by $T_{\rm end}\sim \Lambda_{\rm QCD}$, 
leading to
$\sfrac{T_{\rm end}}{T_{\rm infl}}\sim 10^{-3-4}$,
which corresponds to TeV-scale DM. In this way the DM mass gets connected to approximately the weak scale.

\subsection{The baryon asymmetry}
The baryon asymmetry is washed out  by the super-cooling factor $e^{3N}$.
Thereby the scenario needs to be supplemented by some mechanism that regenerates the baryon asymmetry around the weak scale,
after super-cooling.
This can happen, provided that the three  Sakharov conditions for baryogenesis are satisfied.\footnote{See~\cite{1104.4793,1804.07314} for recent discussions.}
\begin{enumerate}
\item First, deviation from thermal equilibrium can be automatically provided
by the end of super-cooling, either through a first-order phase transition or through the QCD-induced tachionic instability of $s,h$.
\item 
Second, violation of baryon number is automatically provided by sphalerons.
\item 
Third, CP violation.  
\end{enumerate}
The contribution from the CKM phase is too small.
One possibility is to add an axion $a$ coupled to gluons as $\alpha_3 a G_{\mu\nu}^a\tilde{G}^a_{\mu\nu}/8\pi f_a$:
the observed baryon asymmetry can be obtained for values of the axion decay constant allowed by data, $f_a \sim 10^{11}\GeV$~\cite{1407.0030}.
However, depending on the axion model, such a large scale risks conflicting with our assumption of a 
dimension-less theory where the weak scale is dynamically-generated~\cite{1303.7244}.
It would be nice if one could generate a large effective $f_a$ from weak-scale loops, as attempted in~\cite{1507.01755}.
It seems easier to devise  scale-invariant models with extended interactions introduced ad hoc to violate CP:
either extra Yukawa couplings or extra scalar quartics.
One possibility is adding a second Higgs doublet 
such that the scalar potential contains one CP-violating phase
that can lead to baryogengesis~\cite{1203.5012}.
However this also contributes to  electric dipoles,
and flavour data agree with the SM with one Higgs doublet.
Furthermore, detailed computations are needed to establish if the
phase transition predicted by the model is enough out of equilibrium.

In the absence of a lepton asymmetry,
in the above context the reheating temperature must remain below the decoupling temperature 
of electroweak sphalerons, $T_{\rm sph} \approx \unit[132]{GeV}$,
otherwise sphalerons reach thermal equilibrium and wash-out the baryon asymmetry.

Given that observed neutrino masses anyhow demand an extension of the Standard Model,
an appealing alternative possibility developed in the following is low-scale leptogenesis, 
where new neutrino physics generates a lepton asymmetry, converted by 
sphalerons  into the desired baryon asymmetry.






\section{Model with $\SU(2)_X$ gauge group}\label{specific}
The considerations above are fully relevant for basically any dimensionless model that contains a DM candidate (see e.g.~\cite{Hambye:2007vf, Espinosa:2008kw, Foot:2010av, Foot:2011et, Ishiwata:2011aa, Goudelis:2013uca, Steele:2013fka, Carone:2013wla, Farzinnia:2013pga,  Khoze:2013uia, Gabrielli:2013hma, Guo:2014bha, Radovcic:2014rea, Khoze:2014xha, Farzinnia:2014xia, Pelaggi:2014wba, Farzinnia:2014yqa, Altmannshofer:2014vra, Benic:2014aga, Guo:2015lxa, Kang:2015aqa, Endo:2015nba, Plascencia:2015xwa, Ahriche:2015loa, Karam:2015jta, Khoze:2016zfi, Karam:2016rsz, Heikinheimo:2017ofk
}).
Here we consider the model of~\cite{Hambye:2013sna} where
the SM gauge group is extended adding an extra $\SU(2)_X$ with gauge coupling $g_X$, 
and the field content is extended adding a scalar $S$,
doublet  under the extra $\SU(2)_X$, neutral under the SM gauge group.
The Yukawa interactions are those of the SM.
The theory is assumed to be dimension-less, such that the tree-level scalar potential is 
\beq 
V = \lambda_H |H|^4 -\lambda_{HS} |H S|^2 + \lambda_S |S|^4.\label{eq:V}\eeq
This model generates the weak scale through Coleman-Weinberg
dynamical symmetry breaking: the scalar doublets acquire vacuum expectation values and
can be written as
\beq S = \frac{1}{\sqrt{2}} \begin{pmatrix}
0\cr
s
\end{pmatrix},\qquad
H = \frac{1}{\sqrt{2}}
\begin{pmatrix}
0\cr h
\end{pmatrix}\eeq
without loss of generality. 
The Coleman-Weinberg mechanism takes place because
the quartic $\lambda_S$ runs as 
\begin{equation}
\beta_{\lambda_S} \equiv \frac{d\lambda _S}{d\ln\mu} = \frac{1}{(4\pi)^2}\bigg[
\frac{9 g_X^4}{8}-9 g_X^2 \lambda _{{S}}+2 \lambda _{{HS}}^2+24 \lambda _{{S}}^2 
\bigg]\approx  \frac{1}{(4 \pi)^2} \frac{9 \,g_X^4}{8}
\end{equation}
becoming negative at low energy below some scale $s_*$, such that the one-loop potential is approximated as
\begin{equation}
V_1(s) \approx \beta_{\lambda_S}  \, \frac{s^4}{4} \, \ln \frac{s}{s_*}
\end{equation}
which has a minimum at $\med{s} =w= s_* e^{-1/4}$.
The $\SU(2)_X$ vectors acquire a mass $M_X = g_X w/2$, and are stable DM candidates.
Thereby DM has $g_{\rm DM}=9$ degrees of freedom, including the components of
$S$ `eaten' by the massive vectors in the broken phase.
The model {has only two free parameters beyond the ones of the SM: we will use} $g_X$ and $M_X$ as  free parameters.

\smallskip

We compute the other masses assuming, for simplicity,  that
$\lambda_{HS}$ is positive and small.
Then  $\med{s}=w$ induces a Higgs vev
$\med{h} = v$ equal to
$v/w= \sqrt{\lambda_{HS}/2\lambda_H}$,
where $\lambda_H \approx 0.126$ is the SM Higgs quartic, up to small corrections.
This fixes the value of $\lambda_{HS}$ needed to reproduce the desired EW vacuum.
The $s$ mass is $M_s = w \sqrt{\beta_{\lambda_S}}$.
Assuming $M_s \ll M_h$, the $s/h$ mixing angle is
$\alpha\simeq -v/w$.
Finally, a dimension-full constant $V_\Lambda \approx \beta_{\lambda_S}w^4/16
\approx 9M_X^4/8(4\pi)^2$
must be added to the potential such that the true vacuum at $s=w$ has zero energy.  
This is the usual tuning of the cosmological constant.

\subsection{Super-cooling}
At finite temperature the potential receives thermal corrections $V_T$, dominated by
\begin{equation}
V_T(s) =  \frac{9T^4}{2\pi^2} f(\frac{M_X}{T})
+\frac{T}{4\pi}[M_X^3 - (M_X^2 + \Pi_X)^{3/2}]\label{eq:VT}
\end{equation}
where $f(r) =\int_0^\infty x^2 \ln(1-e^{-\sqrt{x^2+r^2}}) dx$ and
$\Pi_X = 5 g_X^2 T^2/6$ is the thermal propagator
for the longitudinal $X$ component
which accounts for re-summation of higher order
ring-diagrams~\cite{hep-ph/0701145,Espinosa:2008kw}.
At small field values, the potential 
is approximated by positive thermal masses for the scalars $s$ and $h$  
\beq
M_s^{2T} = \frac{3}{16} g_X^2 T^2,\qquad
M_h^{2T} =\bigg(\frac{3}{16} g^2_2 + \frac{1}{16}
g^2_Y + \frac{1}{4} y^2_t + \frac12 \lambda_H\bigg)T^2 ,
\label{thermalmass}
\eeq
such that the thermal vacuum is $\med{s}=\med{h}=0$.
As the Universe cools down, a deeper true vacuum appears 
below a critical temperature $T_{\rm cr}$, equal to  $0.31 M_X$ if $g_X \circa{<}0.7$ such that ring diagrams can be neglected,
and roughly a factor $g_X/0.7$ larger otherwise.
Given that dimension-less theories only have thermal masses and quartics,
the Universe remains trapped in the false vacuum at $s=h=0$ down to some temperature $T_{\rm end}$.

Thermal inflation  begins at the temperature  $T_{\rm infl}$ at which the vacuum energy starts dominating with respect to radiation. Applying eq.\eq{TinflH} to our model gives
\begin{equation}
T_{\rm infl}= \left(\frac{135}{64 g_*}\right)^{1/4} \frac{M_X}{\pi} \approx \frac{M_X}{8.5},\qquad
H= \sqrt{\frac{3}{\pi}} \frac{M_X^2}{4M_{\rm Pl}}.
\end{equation}


\subsection{End of super-cooling}
\begin{figure}
\centering
\includegraphics[width=0.45\textwidth,height=0.35\textwidth]{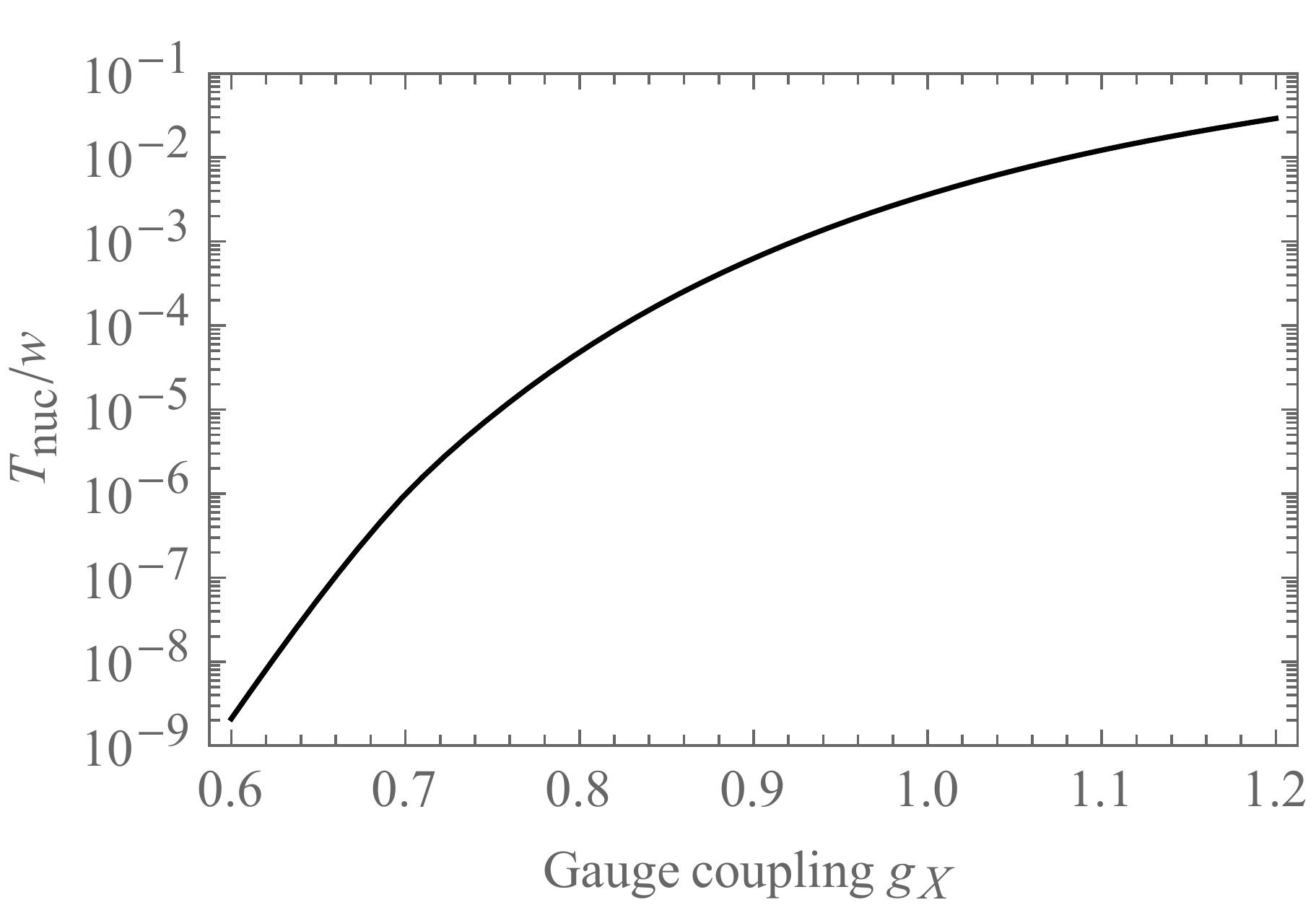}\qquad
\includegraphics[width=0.45\textwidth,height=0.35\textwidth]{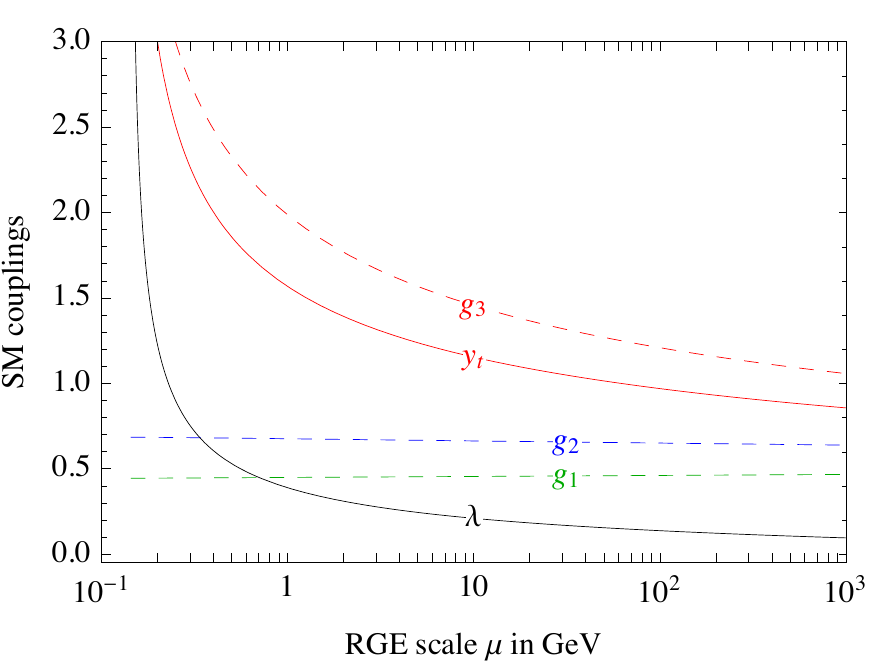}
\caption{\em {\bf Left}:
The nucleation temperature given by vacuum decay, ignoring the QCD phase transition. 
\label{fig:Tnucl}
\label{fig:SMlowRGE} {\bf Right:} 3-loop RGE running in the massless SM.}
\end{figure}

During super-cooling, $s$ and $h$ are kept to 0 by thermal masses.
The temperature  $T_{\rm end}$ at which  thermal inflation ends has the form anticipated in eq.\eq{Tend}.
First, we consider the possibility  that thermal inflation ends through nucleation
and compute $T_{\rm nuc}$.
We solve numerically  the bounce equation\footnote{We assume that the thermal bounce is time-independent and O(3)-symmetric, although this might fail
for dimension-less potentials~\cite{1706.00792}.}
\begin{equation}
s''(r) + \frac{2}{r} s'(r) = \frac{dV}{d s} \;, \qquad s'(0)=0 \; , \quad \lim_{r \to \infty} s(r) = 0
\end{equation}
and use it to calculate the thermal bounce action
\begin{equation}
\frac{S_3(T)}{T} = \frac{4 \pi}{T} \int dr \, r^2 \left[ \frac{1}{2} s'(r)^2 \, +\, V(s(r))\right].
\end{equation}
At $T\ll w$ the potential can be approximated as $ \frac12 M_s^{2T} s^2 +\frac14 \lambda_S(T) s^4$ 
where $\lambda(T)<0$ is the quartic coupling renormalized at $T$, 
and the bounce action as $S_3/T \approx 6.0 \pi M_{s}^T/T|\lambda_S| \approx 8.2 g_X/|\lambda_S|$~\cite{Witten:1980ez}.
Nucleation happens at the temperature $ T_{\rm nuc}$ where
the tunnelling rate is comparable to the Hubble rate,
$S_3(T_{\rm nuc})/T_{\rm nuc} \approx  4\ln M_{\rm Pl}/M_X\approx 142$.  
The numerical results are shown in fig.~\ref{fig:Tnucl}: $T_{\rm nuc}$ 
is very small for small $g_X$.

In such a case, QCD stops super-cooling earlier~\cite{Witten:1980ez,Iso:2017uuu}. 
In the ordinary QCD chiral phase transition scenario where quarks are massive, this phase transition happens at $T_{\rm cr}^{\rm QCD}\approx  \unit[154\pm9]{MeV}$~\cite{1203.5320}.
However, during super-cooling all quarks are massless, which leads to a smaller value of  $\alpha_3$ at low energy.
Fig.\fig{SMlowRGE} shows the running of the SM couplings: 
$\alpha_3(\bar\mu)$ diverges at $\Lambda_{\rm QCD}^{h=0}\approx 144\MeV$,
with $\Lambda^{(6)}_{\overline{\rm MS}} = (89\pm7)\MeV$ if only
 $\alpha_3$ is kept in the RGE~\cite{pdg}.
Then  the QCD chiral phase transition happens at  a lower temperature, $T_{\rm cr}^{\rm QCD}\sim \unit[85]{MeV}$
according to the estimate of~\cite{hep-ph/0602226,Iso:2017uuu}.
When a zero-mode quark condensate forms, the
Yukawa coupling $y_t h \med{t_L t_R}/\sqrt{2} + \hbox{h.c.}$
induces a linear term in the Higgs potential,
such that the Higgs  acquires a $T$-dependent vacuum expectation value 
$\med{h}_{\rm QCD}$.
Given that the couplings $y_t$ and $\lambda_H$  too run to non-perturbative values (see fig.\fig{SMlowRGE})
 $\med{h}_{\rm QCD}$ can at best be estimated.
We will proceed assuming $\med{h}_{\rm QCD}\approx \unit[100]{MeV}$,
up to order one factors.



\smallskip

Next, $\med{h}$  induces a mass term for the $s$ scalar, 
$M_s^2 =  - \lambda_{HS} \med{h}^2/2$.
If $\lambda_{HS}<0$, the positive $M_s^2$ delays the end of thermal inflation. 
If $\lambda_{HS}$ is positive ({as needed to break ${\rm SU(2)}_L$ at the true minimum})
the negative $M_s^2$ triggers the end of thermal inflation:
$s$ too starts rolling down as soon as its extra mass term $M_s$ becomes larger than its thermal mass $M_s^{T}$
in eq.~(\ref{thermalmass}).
If $\lambda_{HS}$ is large enough, this happens immediately at $T_{\rm end}=T_{\rm cr}^{\rm QCD}$;
otherwise this happens later at a lower temperature
\beq\label{eq:endQCD}
T_{\rm end}^{\rm QCD}
=\sqrt{\frac{8 \lambda_{HS}}{3}}  \frac{ \med{h}_{\rm QCD}}{g_X} \approx
\frac{0.1 \, \langle h \rangle_{\rm QCD} }{M_X/\unit{TeV}}.
\eeq
In the present model the thermal mass $M_s^T$ is dominated by DM vectors, 
so that thermal inflation ends when their density is diluted enough.
%
%
%
%

\begin{figure}
\centering
\includegraphics[width=0.423\textwidth]{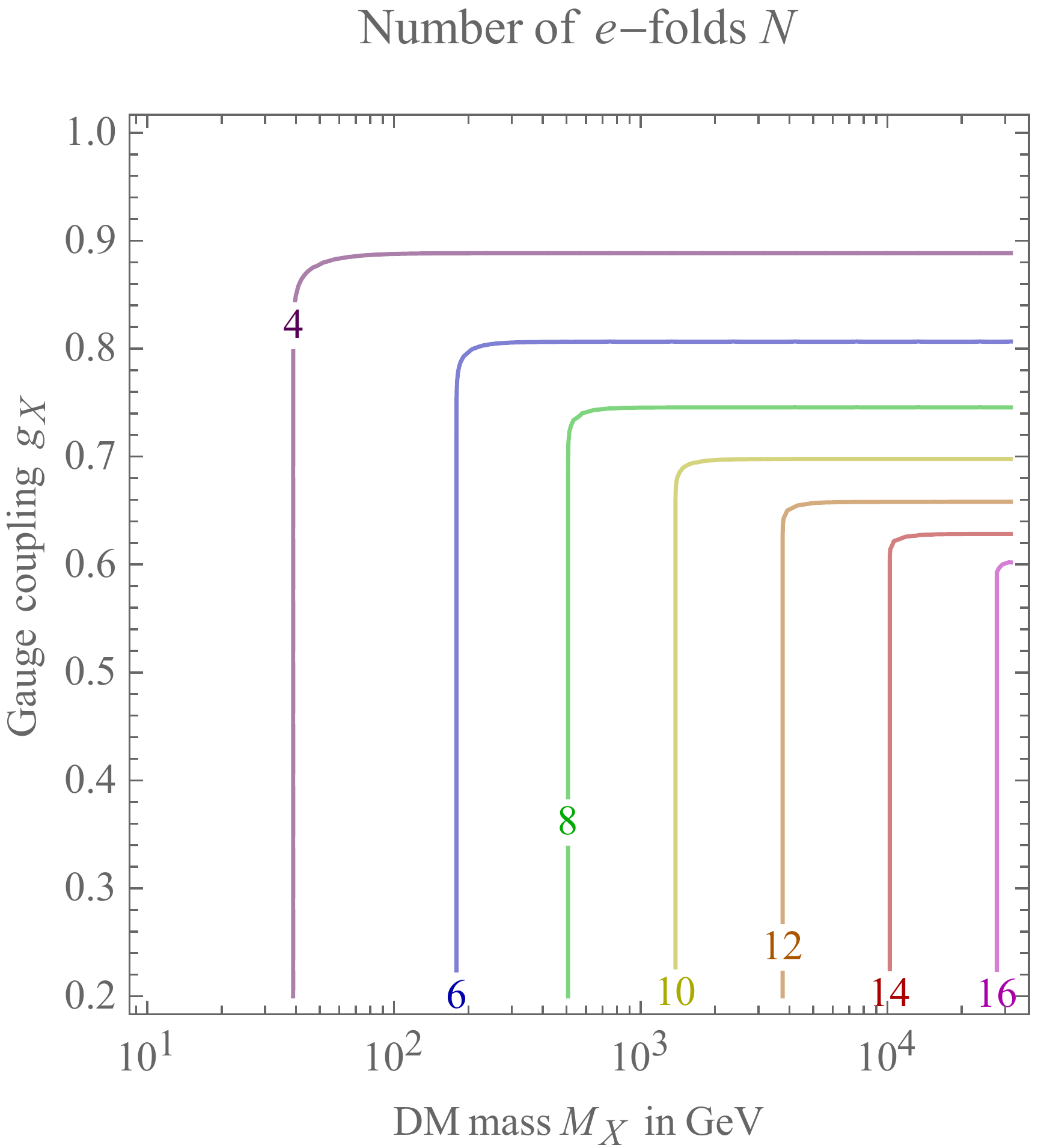}\qquad
\includegraphics[width=0.45\textwidth]{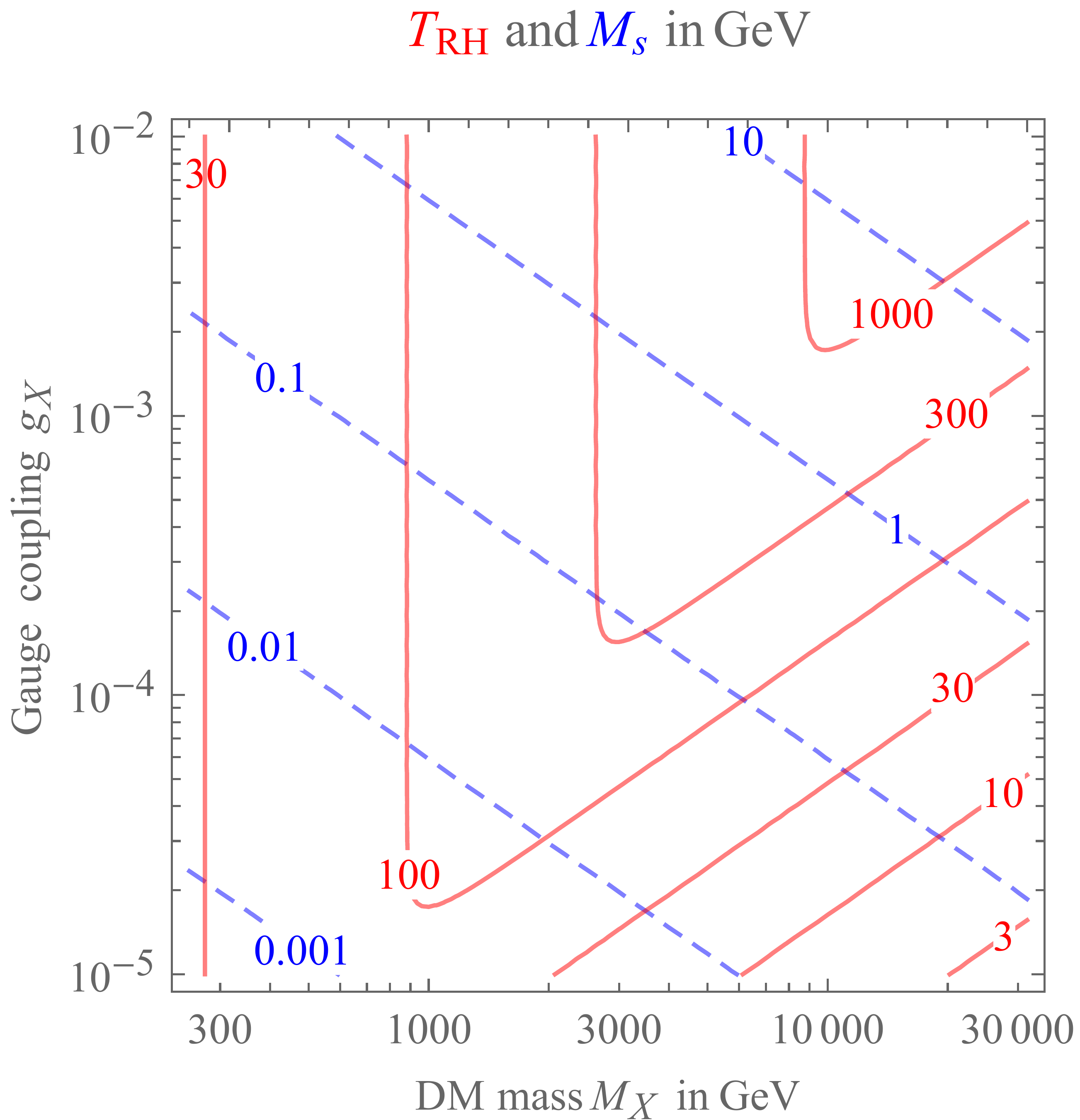}
\caption{\em {\bf Left:} 
number of $e$-folds of thermal inflation, $N = \ln T_{\rm infl}/T_{\rm end}$ for $\langle h \rangle_{\rm QCD} =  \unit[100]{MeV}$.
{\bf Right:}   reheating temperature in $\GeV$ (solid red curves)
and $M_s/\GeV$ (diagonal dashed lines).\label{fig:reheating_m_g}\label{fig:infl}}
\end{figure}

We now have all the factors that determine $T_{\rm end}$  in eq.\eq{Tend}.
During  super-cooling, the Universe inflates by a factor $T_{\rm infl}/T_{\rm end}$ plotted in fig.~\ref{fig:infl}a. The horizontal part of the contours corresponds to end of super-cooling via vacuum decay, and the vertical part to the QCD-triggered end.

\subsection{Reheating}
After the end of inflation, the scalars oscillate around the true minimum, 
dissipating their energy density $\rho_{\rm sca}$ with some rate $\Gamma$
into radiation that acquires energy density $\rho_{\rm rad}= \sfrac{g_* \pi^2 T^4}{30}$. 
The rolling fields $s$ and $h$ finally settle at the true minimum.
The scale factor $a$ and the various components evolve as
\beq\label{eq:Bolz}
\left\{ \begin{array}{rcl}
\displaystyle
\frac{\dot a}{a} &=& H = \displaystyle
 \frac{1}{\bar M_{\rm Pl}} \sqrt{\frac{\rho_{\rm sca} + \rho_{\rm rad}}{3}}\\
\dot\rho_{\rm sca} &=& - (3H + \Gamma) \rho_{\rm sca}  \\
\dot \rho_{\rm rad} &=& -4 H \rho_{\rm rad} + \Gamma \rho_{\rm sca}\\
\dot n_{\rm DM} &=& - 3H n_{\rm DM}  +
 \langle \sigma v \rangle_{\rm ann} (  n_{\rm DM}^{\rm eq2}-n^2_{\rm DM}  ) +
 \langle \sigma v \rangle_{\rm semi} n_{\rm DM}(  n_{\rm DM}^{\rm eq} -n_{\rm DM} )  .
\end{array}\right.\eeq
Thereby $ \rho_{\rm sca} (t)= \rho_{\rm sca}(t_{\rm end}) {e^{-\Gamma(t-t_{\rm end})} [a(t_{\rm end})/a(t)]^3}$.
This roughly means that the inflaton $s$ reheats the Universe up to the temperature
\beq
T_{\rm RH} =  \left( \frac{45}{4\pi^3 g_*}\right)^{1/4}   M_{\rm Pl}^{1/2}  \min(H, \Gamma)^{1/2}
=T_{\rm infl} \min (1, \frac{\Gamma}{H})^{1/2}.
\label{eq:TRH}
\eeq
We need to compute $\Gamma$.
The scalar equations of motion are
\beq
\left\{\begin{array}{rcl}
\ddot s + (3H+\Gamma_s) \dot s \!\!&=&\!\!- {\partial V}/{\partial s} \cr
 \ddot h + (3H+\Gamma_h) \dot h \!\!&=&\!\!- {\partial V}/{\partial h} 
\end{array}\right.\qquad\Rightarrow\qquad
\frac{d}{dt} (K+V)= - 6 H K- \Gamma_h {K_h} - \Gamma_s {K_s}
\eeq
where $K =K_h+K_s=(\dot s^2 + \dot h^2)/2$ is the scalar kinetic energy.
Given that scalar masses are much bigger than $H$,
averaging over the fast oscillations around the minimum,
one finds that the scalar energy
$\rho_{\rm sca}=\med{K+V} = 2\med{K}$
red-shifts as non relativistic matter, in the limit  $\Gamma_h=\Gamma_s=0$.

%

All masses stay 
positive around the true minimum, so decays are not enhanced by parametric resonances. 
Most of the energy is stored in $s$, but its decay rate $\Gamma_s$ can be smaller than $H$.
On the other hand the Higgs potential energy is sub-leading, while its decay rate $\Gamma_h \approx 4\MeV$ is fast.
Thereby the decay rate $\Gamma$ of the combined system is controlled by 
the rate for energy transfer from $s$ to $h$
due to the $\lambda_{HS}$ interaction.
We compute $\Gamma$ by solving the equations of motion in linear approximation around the
minimum
\begin{align}
\ddot{h} + (3H+\Gamma_h) \dot{h} &= -\tilde{M}_h^2  (h-v) \,+\, \lambda_{HS} v w (s-w) \\
\ddot{s}+(3H+\Gamma_s) \dot s &= -\tilde{M}_s^2  (s-w) \,+\, \lambda_{HS} v w (h-v) .
\end{align}
where $\tilde{M}_h^2 = 2 \lambda_H v^2$, $\tilde{M}^2_s = \beta_{\lambda_S} w^2$.
The complex frequencies of the normal modes of damped oscillations
in the limit $\Gamma_{s}, H\ll \Gamma_h \ll M_h$ are 
$\omega \simeq M_h \pm i \sfrac{(3H+\Gamma_{h})}{2}$ and
$\omega \simeq M_s \pm i \sfrac{(3H+\Gamma)}{2}$ with
\beq \Gamma = \Gamma_h \, \sin^2 \alpha  + \Gamma_s \cos^2\alpha \eeq
where $\alpha \simeq - v/w =\sqrt{\lambda_{HS}/2\lambda_H} \ll 1$ is the angle that diagonalizes the mass matrix.

Fig.~\ref{fig:reheating_m_g}b shows the numerical results for $T_{\rm RH}$.
For a given $M_X$, reheating is instantaneous provided that $g_X$ is large enough, 
so that $T_{\rm RH} = T_{\rm infl} \simeq M_X/8.5$, corresponding to  vertical
contour lines in fig.~\ref{fig:reheating_m_g}b. 
For smaller $g_X$ the reheating temperature is suppressed.\footnote{For very small $g_X$ one can have $T_{\rm RH} \lesssim \unit[100]{MeV}$ so that macroscopic six-flavour quark nuggets~\cite{1804.10249} could contribute to the DM relic density.}

\begin{figure}[t]
$$\includegraphics[width=0.47\textwidth,height=0.45\textwidth]{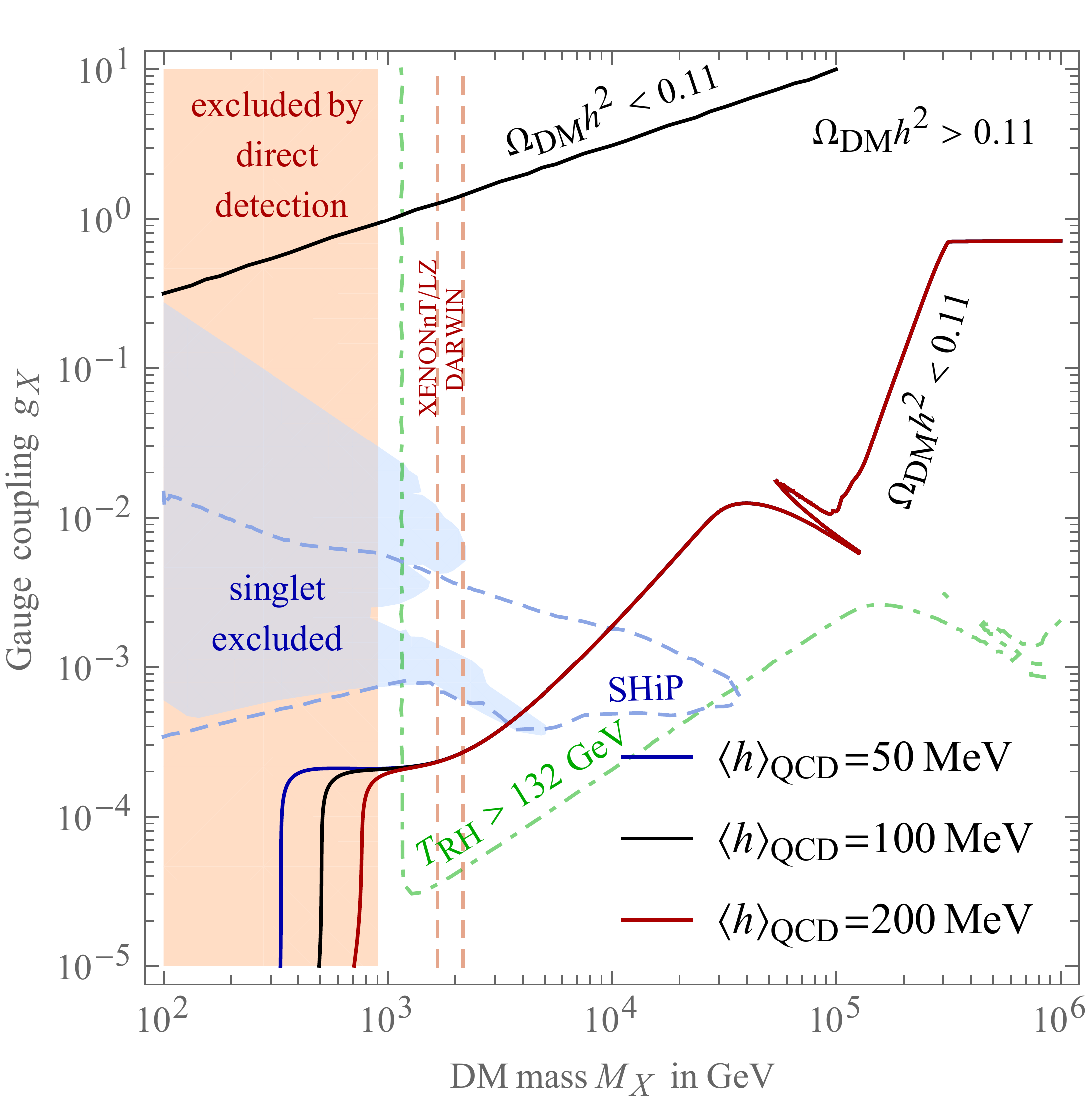}\qquad
\includegraphics[width=0.47\textwidth,height=0.47\textwidth]{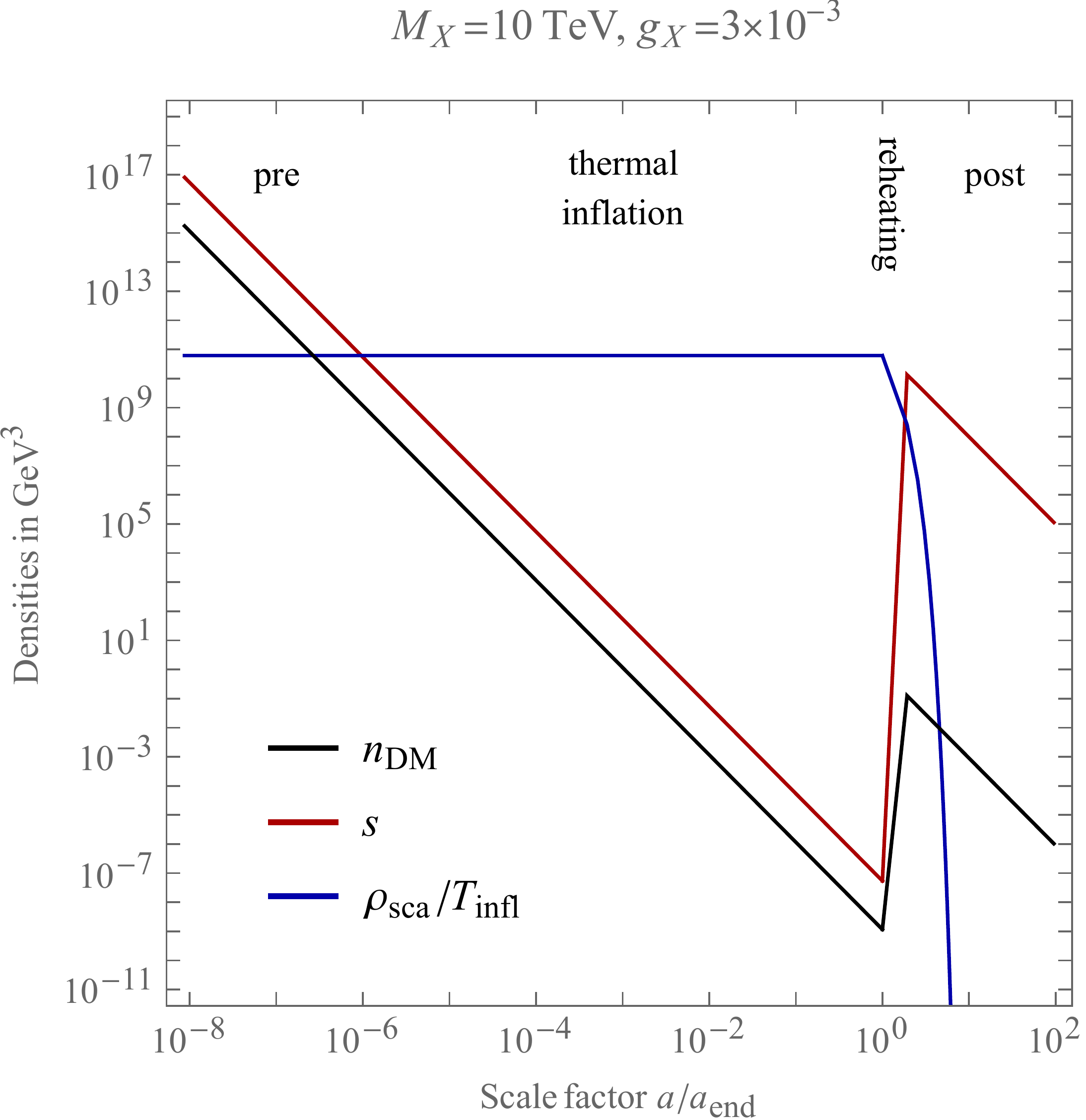}$$
\caption{\em  {\bf Left:}
The observed DM abundance is reproduced along the solid curves, computed for different values
of the uncertain QCD factor $\langle h \rangle_{\rm QCD}$.
The region shaded in orange (blue) is excluded by direct DM searches
(collider searches for the singlet $s$).
Dashed curves show future detection prospects. \label{fig:m_g}
{\bf Right:}
Sample evolution of the DM density, of entropy, of the scalar energy. \label{fig:evo}}
\end{figure}

\subsection{The dark matter abundance}
The DM candidates are the $\SU(2)_X$ vectors with mass $M_X$.
In the usual scenario where they are thermal relics, the
observed DM abundance is reproduced for $g_X^2 \approx M_X/\TeV$~\cite{Hambye:2013sna}
and super-cooling is negligible.
Smaller values of $g_X$ lead to super-cooling, realising the novel DM production mechanism proposed here:
the DM abundance is the sum of the super-cool population, plus the sub-thermal population, Eq.~(\ref{twopopulations}).

We compute the super-cool DM population specializing eq.~\eqref{eq:0.40eV}
and eq.~\eqref{eq:Ysc} to the present model.
We find that the super-cool abundance reproduces the observed DM abundance when
the end of super-cooling is triggered by the QCD phase transition
as $T_{\rm end} = T_{\rm end}^{\rm QCD}$ (eq.~\eqref{eq:endQCD}),
and when reheating is instantaneous.
Thereby $T_{\rm RH} = T_{\rm infl} \approx M_X/8.5$ and the DM abundance simplifies to
\begin{equation}
\Omega_{\rm DM} h^2|_{\rm super-cool}
\approx 3.7\times 10^{-3} \, \left( \frac{\langle h \rangle_{\rm QCD}}{\unit[100]{MeV}}\right)^3 \left( \frac{\unit{TeV}}{M_X} \right)^5.
\end{equation}
It does not depend on $g_X$, giving rise to the vertical contours in the $(M_X, g_X)$ plane in the left panel of fig.\fig{m_g}.

As anticipated, this is not the end of the story:
one needs to take into account the effects of thermal scatterings after reheating.
One needs to evolve the Boltzmann equation in eq.~\eqref{eq:Bolz} starting from
the initial condition $Y_{\rm DM}(T_{\rm RH}) = Y_{\rm DM}|_{\rm super-cool}$.
The $s$-wave cross-sections for DM annihilations
$VV\leftrightarrow ss$ and  semi-annihilations $VV\leftrightarrow V s$ are~\cite{Hambye:2013sna}
\beq \langle \sigma v\rangle_{\rm ann} =\frac{11g_X^4}{6912\pi M_X^2},\qquad
\langle\sigma v\rangle_{\rm semi-ann} = \frac{g_X^4}{128\pi M_X^2}.
\eeq
In the extreme case where the reheating temperature is larger than the DM decoupling temperature, the 
super-cool population is erased and substituted by the usual thermal relic population.
Otherwise, the super-cool population is negligibly suppressed, and complemented by the additional sub-thermal population
of eq.\eq{YDMthermal}.
For instantaneous reheating ($z_{\rm RH}\approx 8.4$) this evaluates to
$\Omega_{\rm DM} h^2|_{\rm sub-thermal}\approx 0.110 (g_X/0.00020)^4$,
giving rise to the horizontal part of the contour at $g_X \approx 10^{-4}$ in the  $(M_X, g_X)$ plane of fig.\fig{m_g}.
At larger $M_X$ reheating is no longer instantaneous, giving rise to the oblique part of the contour in
fig.~\ref{fig:m_g}a, which shows the complete numerical results for the DM density. Finally, for even larger masses $M_X \gtrsim \unit[300]{TeV}$ the super-cool abundance reproduces the observed DM relic density with $g_X \sim O(1)$, so that super-cooling is ended by nucleation, leading to a DM abundance that depends mainly on $g_X$ and not on $M_X$. Therefore, DM masses can be even PeV-scale or larger, higher than those allowed with freeze-out and perturbative couplings. At $M_X \approx \unit[100]{TeV}$ one has $M_s \simeq M_h$, so that a resonance-like feature in the mixing angle $\alpha$, and consequently in $T_{RH}$ and $\Omega_{DM}$, appears.

In the  region of the parameter space relevant for the present work,
the Spin-Independent cross section for DM direct detection
is dominantly mediated by 
$s$ and simplifies to
\beq \sigma_{\rm SI} = \frac{m_N^4  f^2 }{16\pi v^2}
 \bigg(\frac{1}{m_s^2} - \frac{1}{m_h^2}\bigg)^2 g_X^2 \sin^2 2\alpha
 \simeq  \frac{64\pi^3 f^2  m_N^4}{81 M_X^6}\approx
 0.6~10^{-45}\cm^2  (\frac{\TeV}{M_X})^6
\eeq
where $f\approx 0.295$ is the nucleon matrix element and $m_N$ is the nucleon mass. 
So $\sigma_{\rm SI} < 1.5~10^{-45}\cm^2 (M_X/\TeV)$~\cite{1705.06655} for
$M_X > 0.88\TeV $.

Fig.~\ref{fig:m_g}a  also shows the existing bounds and future discovery prospects, coming both from searches of dark matter via direct detection, and from collider searches for $s$.
The region shaded in orange is excluded by direct searches at XENON1T~\cite{1705.06655} and the dashed vertical lines denote the future sensitivity of XENONnT~\cite{1512.07501}, LZ~\cite{1509.02910} and DARWIN~\cite{1606.07001}.
The direct detection cross section is suppressed by two powers  of the small $g_X$ coupling
and enhanced by the exchange of the 
$s$ scalar state, which is light, see fig.~\ref{fig:infl}b.
As a result the direct detection constraint is significant.
The region shaded in blue is excluded by collider searches for $s$:
the dominant collider bounds come from $s \to e e, \mu \mu$ at CHARM and $B \to K^* \mu \mu$ at LHCb,
as summarized in~\cite{SHiP-NOTE}.
The dashed blue curves indicates the future sensitivity of SHiP~\cite{Anelli:2015pba,SHiP-NOTE}. 

\medskip

Finally,  fig.~\ref{fig:m_g}b shows a sample example of the evolution of the DM density,
of entropy, of the energy density in scalars:
the latter dominate during super-cooling, while $n_{\rm DM}$ and $s$ decrease equally.
At reheating almost all of this energy is transferred to entropy, and only a small fraction goes to massive DM.

\subsection{The baryon asymmetry}\label{Yb}
The reheating temperature $T_{\rm RH}$  in the $\SU(2)_X$ model 
is shown in fig.~\ref{fig:infl}b and can be either smaller or larger  than $T_{\rm sph}$, see fig.\fig{m_g}a.
In the first case cold baryogenesis might be a viable option (with extra CP-violation),
while leptogenesis is a clear option in the second case.

Leptogenesis can in particular be achieved if one adds right-handed neutrinos $N$, with Yukawa couplings $Y_N\, NLH$,
and an extra real scalar singlet $S'$, with quartic potential couplings and a
Yukawa coupling $y_S\,S' N^2/2$ that breaks lepton number. This 
 induces a mass $M_N$ for $N$ if $S'$ acquires a vev.

Low-scale leptogenesis with right-handed neutrinos and their Yukawa couplings 
can then occur either: 
\begin{enumerate}
\item[1)] via resonant CP-violating decays of the right-handed neutrinos~\cite{Pilaftsis:1997jf,Pilaftsis:2003gt,Dev:2014laa};
\item[2)] via oscillations of right-handed neutrinos with total $L$ lepton number conservation (ARS framework~\cite{Akhmedov:1998qx,Asaka:2005pn});
\item[3)] via $L$-violating Higgs decays~\cite{Hambye:2016sby,Hambye:2017elz}. 
\end{enumerate}
Barring fine-tunings in the structure of Yukawa couplings, a sizeable lepton asymmetry needs quasi-degenerate right-handed neutrinos, at the per-million level.
The time-variation of their masses (while scalars relax to their minimum) relaxes the amount of
quasi-degeneracy, a scenario we will not further explore here.
%

\begin{figure}[t]
\centering
\includegraphics[width=0.5\textwidth]{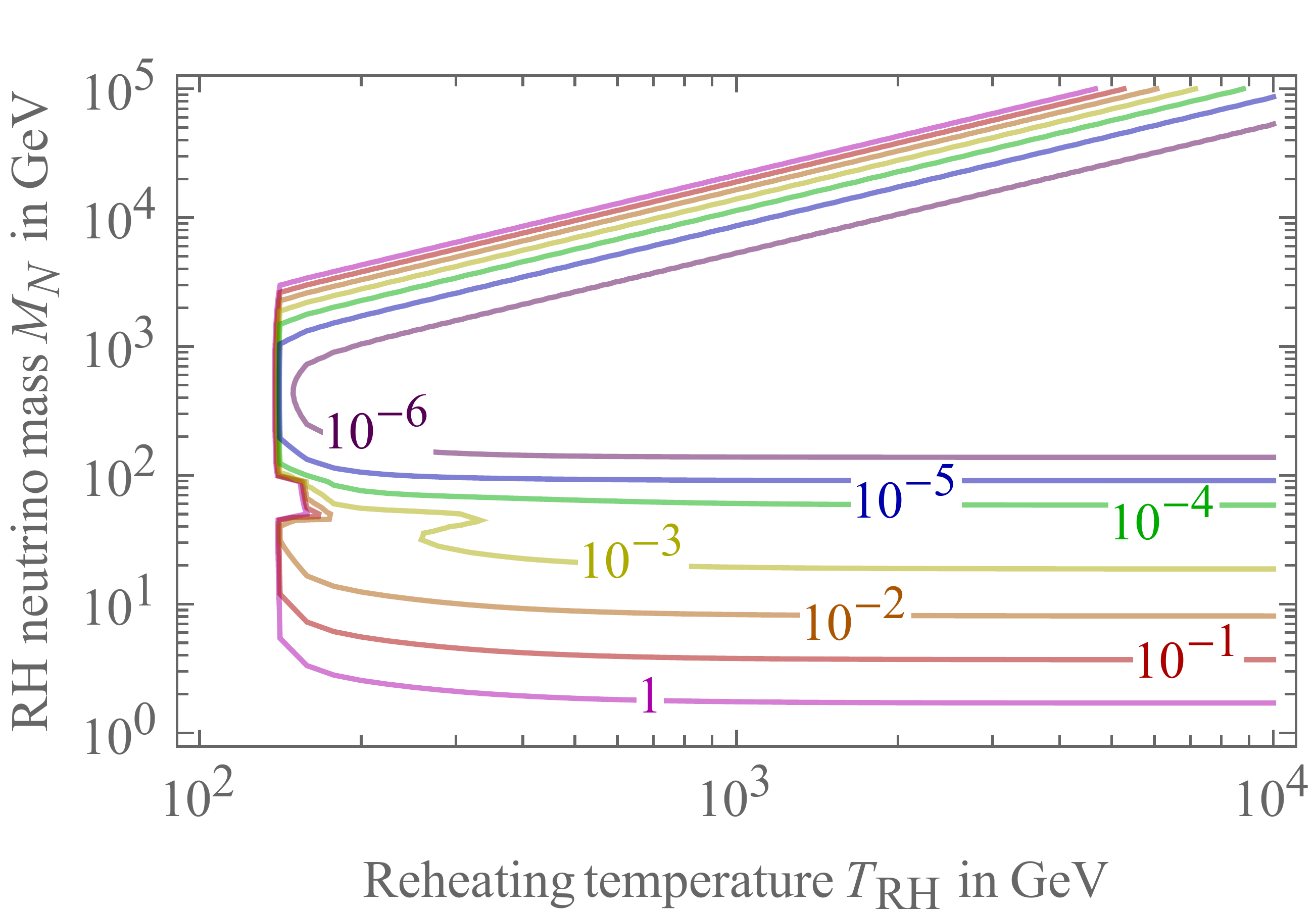}
\caption{\em\label{fig:lepto_TRH}  
Leptonic {\rm CP} asymmetry $\epsilon_{\rm CP}$ needed to obtain successful leptogenesis, assuming
right-handed neutrinos in thermal equilibrium at the temperature $T_{\rm RH}$. 
For $M_N \gtrsim M_h$ ($M_N \circa{<} M_h$)
the asymmetry comes from decays of right-handed neutrinos (of the Higgs),
and the needed $\epsilon_{\rm CP}$ is obtained from
right-handed neutrinos degenerate at the $\Delta M_N/M_N \approx 10^{-7}$ level
(at the $10^{-5}$ level). We fixed the Yukawa couplings to $|Y_N|^2 v^2/M_N = \unit[10^{-11}]{eV}$.}
\end{figure}

The first possibility, resonant CP-violating decays, 
mostly produces an asymmetry for $T\sim M_N$ and thereby needs $T_{\rm RH} \circa{>} M_N$.
Indeed the right-handed neutrino mass must be sizeably above the electroweak scale to allow $N\to HL$ decays and to produce efficiently the asymmetry before $T_{\rm sph}$.
In the super-cool DM production scenario above, 
the reheating temperature can  be  above a TeV, see fig.~\ref{fig:reheating_m_g}b. 
This longer period of electroweak symmetry breaking restoration makes this scenario easily viable.
This is shown in fig.~\ref{fig:lepto_TRH}. 
Successful leptogenesis implies a lower bound on the reheating temperature, depending on $M_N$,
which implies a lower bound on $M_X$. 

\smallskip

The second possibility of low scale leptogenesis, right-handed neutrino oscillations in $L$-conserving processes, 
requires  lighter $N$, around the GeV scale. 
However, it is in general fully operational at temperatures orders of magnitudes larger than the electroweak scale.
Thus, except in special situations, in our context it is suppressed by the low reheating temperature.

\smallskip

The third possibility, $L$-violating Higgs decay (which requires right-handed neutrino mass
between a GeV up to the Higgs mass),
produces dominantly the baryon asymmetry at temperatures just above the sphaleron decoupling temperature $T_{\rm sph}$.
Therefore, one only needs $T_{\rm RH}\circa{>}T_{\rm sph}$,
which can be realised in the allowed parameter space of fig.~\ref{fig:m_g}.
This mechanism explains why in fig.~\ref{fig:lepto_TRH}, 
which combines the leptogenesis contributions from $L$-violating 
$N$ and $H$ decays,
leptogenesis is viable for masses below the Higgs boson mass and $T_{\rm RH} \geq T_{\rm sph}$.
We solved Boltzmann equations taken from~\cite{Hambye:2016sby} in the single-flavour approximation for the SM leptons. These do not take into account the reheating temperature suppressed purely-flavoured ARS contribution.

In all cases (including the model discussed in section~\ref{B-L} below), leptogenesis is significantly facilitated by the super-cool mechanism, since the required gauge couplings are small and do not dilute the asymmetry as they do, instead, in the WIMP regime~\cite{Frere:2008ct,Heeck:2016oda}, where successful leptogenesis would hardly be possible.


%
%
%

\section{Model with ${\rm U}(1)_{B-L}$ gauge group}\label{B-L}
We here study a different super-cool DM model.
Given that leptogenesis seems the most plausible option for baryogenesis,
a natural possibility is gauging $B-L$, such that  right-handed neutrinos become necessary for anomaly cancellation.
This makes $\SU(2)_X$  no longer necessary for dynamically breaking scale invariance:
the role of $\SU(2)_X$ can be played by U$(1)_{B-L}$, in a similar way.
The scalar doublet $S$ of $\SU(2)_X$ is replaced by a complex scalar, $S$, charged under U$(1)_{B-L}$. 
The gauge coupling $g_{B-L}$ can drive the scalar quartic $\lambda_S$ to run negative around the weak scale,
such that the scalar $S$ again acquires a vacuum expectation value.
As a result, the $B-L$ gauge boson $Z'$ acquires a mass, eating the would-be Majoron.
The weak symmetry is again broken thanks to a $-\lambda_{HS}|SH|^2$ term in the potential with
$\lambda_{HS}>0$.


Assuming that it has a $B-L$ charge equal to 2, the $S$ scalar can be identified with the field
that gives mass to right handed neutrinos $N$, through a $Y_S~SN^2/2$ Yukawa interaction.

A disadvantage of this model  (compared to the $\SU(2)_X$ model)
is that the $Z'$ cannot be DM: it decays into SM fermions, as they are 
charged under $B-L$; furthermore $B-L$ can have a kinetic mixing with hypercharge.
We thereby add one extra scalar singlet $\phi$, with no hypercharge and $B-L$ charge $q_\phi$ chosen such that it is stable:
for simplicity we assume $q_\phi=1$.
This makes DM absolutely stable due to the fact that $\phi$ is odd under the $\mathbb{Z}_2 \subset {\rm U}(1)_{B-L}$ symmetry, 
which remains unbroken because $S$ has $B-L$ charge 2.\footnote{
This  $B-L$ {gauge group} has been considered in its scale invariant version  in~\cite{Iso:2009nw,Iso:2017uuu} 
as a model for neutrino masses (without DM) and in \cite{Khoze:2014xha},
with a DM scalar particle which has no $B-L$ charge (in order to avoid direct detection bounds)
and is stabilised adding an extra $\mathbb{Z}_2$ symmetry.  
Super-cool DM, {instead}, requires small values of the $B-L$ gauge coupling, 
such that we can assume that DM is charged under $B-L$, and thus automatically stable,
compatibly with direct detection constraints.}

Summarising, 
the model is described by gauge-invariant kinetic terms, plus
the dimension-less scalar potential plus a constant
\beq
V=\lambda_H |H|^4  +\lambda_S |S|^4 
+\lambda_\phi  |\phi|^4
-\lambda_{HS} |HS|^2
+\lambda_{S\phi} |S\phi|^2
+\lambda_{H \phi}|H\phi|^2 + V_\Lambda
\eeq
plus the Yukawa interactions
\beq
\Lag_{\rm Yuk}= \Lag_{\rm Yuk}^{\rm SM} +Y_N {N} L H + Y_S \, S \frac{N^2}{2} + \hbox{h.c.}
\eeq
After symmetry breaking 
the scalar fields can be written as
\beq S = \frac{s}{\sqrt{ 2}},\qquad
H = \frac{1}{\sqrt{2}}
\begin{pmatrix}
0\cr h
\end{pmatrix}.\eeq
At one loop, the $\lambda_S$ quartic runs as  
\begin{equation}
\beta_{\lambda_S} \equiv \frac{d\lambda _S}{d\ln\mu} = \frac{1}{(4\pi)^2}\bigg[
96 g_{B-L}^4- Y_S^4 + 2\lambda_{HS}^2+\lambda_{S\phi}^2 + 20 \lambda_S^2 + \lambda_S(2 Y_S^2 - 48 g_{B-L}^2)
\bigg]
\end{equation}
becoming negative at low energy below some scale $s_*$, such that its one-loop potential is approximated as
\begin{equation}
V_1(s) \approx \beta_{\lambda_S}  \, \frac{s^4}{4} \, \ln \frac{s}{s_*}
\end{equation}
which develops a minimum at $\med{s} =w= s_* e^{-1/4}$.
This generates
\beq
M_{Z'}=2g_{B-L} w,\qquad M_N= Y_S\, w
\eeq
as well as electro-weak symmetry breaking and neutrino and DM masses,
\beq
\frac{v}{w}=\sqrt{\frac{\lambda_{HS}}{2\lambda_H}} ,\qquad m_\nu=- Y_N^T \cdot \frac{v^2}{M_N}\cdot   Y_N ,\qquad
M_{\rm DM}=M_\phi =\sqrt{\frac{\lambda_{S\phi}}{2}} w.
\eeq
We neglected the contribution of $\lambda_{H\phi}$ to $M_\phi$.
Electroweak precision data imply the bound $M_{Z'}/g_{B-L} \circa{>} 7\TeV$~\cite{hep-ph/0604111,0909.1320},
up to corrections due to kinetic mixing.
DM has $g_{\rm DM}=2$ degrees of freedom and
is not destabilised by the symmetry breaking of the various gauge groups
provided that $\lambda_{S\phi}, \lambda_{H\phi} ,\lambda_\phi>0$ such that
the DM scalar  $\phi$ does not acquire a vacuum expectation value.
The condition $\lambda_\phi>0$ is easily satisfied in view of the smaller $g_{B-L}^4$ contribution to its running:
\beq \beta_{\lambda_\phi} = \frac{1}{(4\pi)^2}\bigg[6 g_{B-L}^4+2\lambda_{H\phi}^2+\lambda_{S\phi}^2+20\lambda_\phi^2
-12\lambda_\phi g_{B-L}^2\bigg].\eeq
The vacuum energy vanishes for {$V_\Lambda \approx  \beta_{\lambda_S}w^4/16
\approx (3 M_{Z'}^4/8 + M_{\rm DM}^4/4)/(4\pi)^2$} such that 
{$T_{\rm infl} \approx (M_{Z'}^4 + 2 M_{DM}^4/3)^{1/4}/11$}.
The $s$ thermal potential is
{$V_T \approx T^4[3 f(\sfrac{M_{Z'}}{T}) + 2 f(\sfrac{M_{\rm DM}}{T})]/2\pi^2$} and
its thermal  mass is
{$M_s^{2T} = (g_{B-L}^2 + \lambda_{S\phi}/12) T^2$}, so that $T_{\rm end}^{\rm QCD}
=  \med{h}_{\rm QCD} \sqrt{ \lambda_{HS}/(2g_{B-L}^2 + \lambda_{S\phi}/6)}$.

\begin{figure}[t]
\centering
$$\includegraphics[width=0.45\textwidth,height=0.45\textwidth]{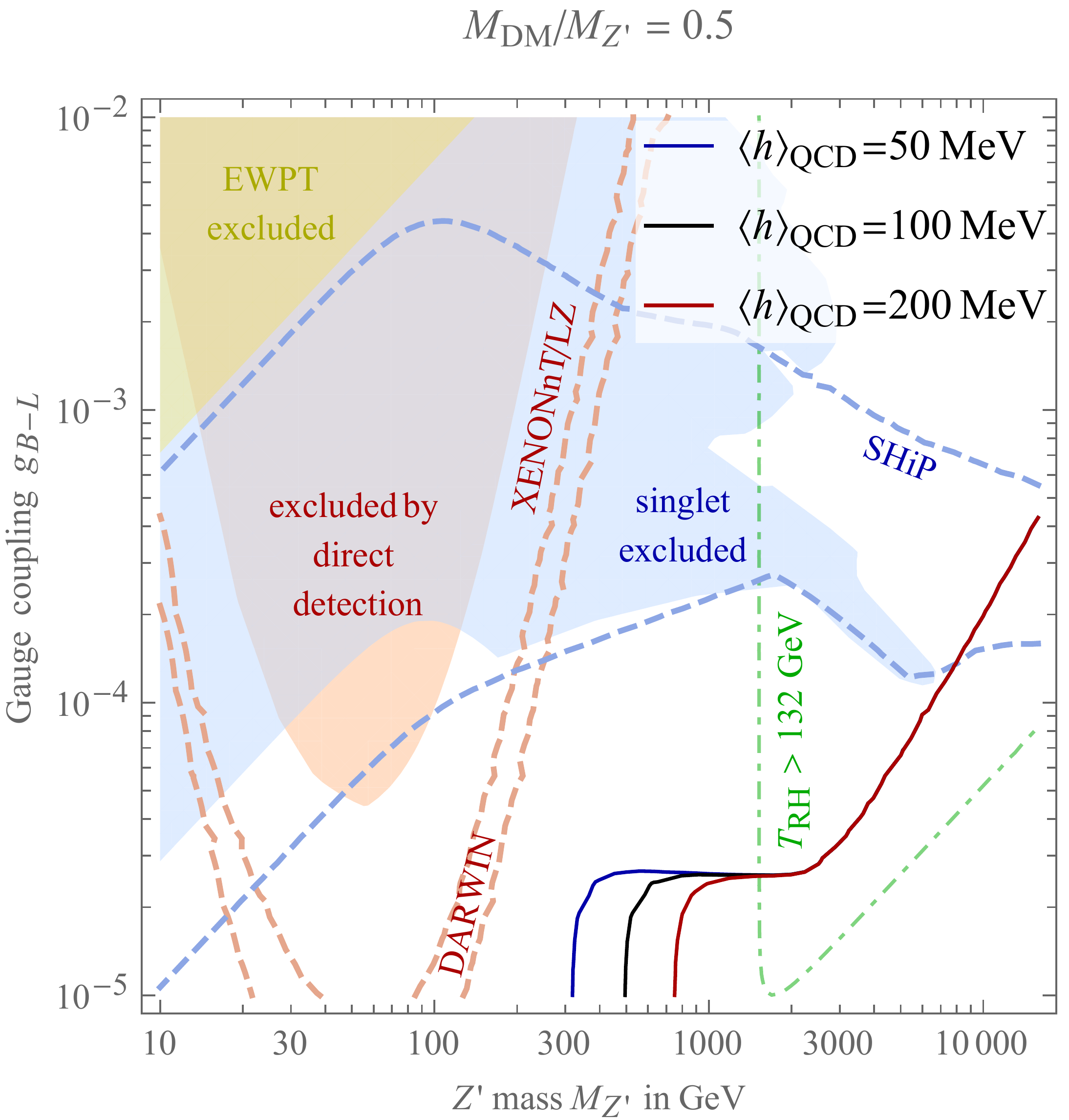}\qquad
\includegraphics[width=0.45\textwidth,height=0.45\textwidth]{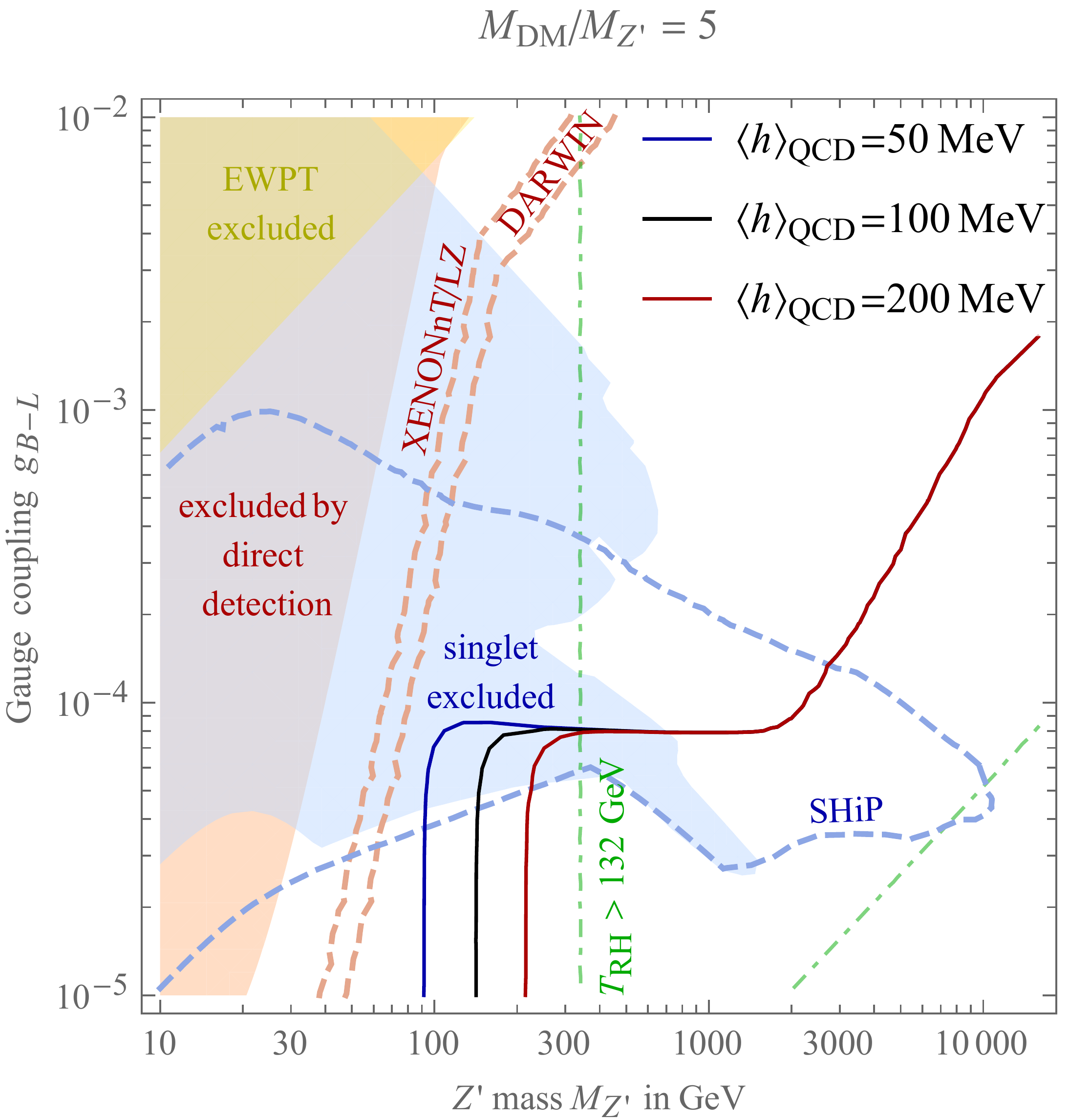}$$
\caption{\em The observed DM abundance is reproduced along the solid curves, 
computed for different values
of the uncertain QCD factor $\langle h \rangle_{\rm QCD}$.
The region shaded in orange (blue) is excluded by direct DM searches
(collider searches for the singlet $s$).
The region shaded in yellow is excluded by precision data.
Dashed curves indicate future detection prospects.\label{fig:BmL}}
\end{figure}

\medskip

The Spin-Independent cross section for DM direct detection receives two unavoidable contributions,
from $Z'$ mediation, and from $\lambda_{S\phi}$ (via the small
mixing $\alpha \simeq -v/w$ between $h$ and $s$),
as well as a contribution from $\lambda_{H\phi}$:
\beq \sigma_{\rm SI} \approx \max\bigg(\frac{g_{B-L}^4  q_\phi^2 m_N^2}{4 \pi M_{Z'}^4},
\frac{\lambda_{S\phi}^2 m_N^4 f^2 \sin^2 2\alpha}{16\pi M_{\rm DM}^2 M_s^4},
\frac{\lambda_{H\phi}^2 m_N^4 f^2 
}{16\pi M_{\rm DM}^2 M_h^4}
\bigg).\eeq
Finally, we need the DM {pair production} cross section that {is at the origin of} the sub-thermal population.
In the non-relativistic limit, {DM is produced in pairs in a $s$ wave way from two $Z'$, from two scalars
(possibly through a $Z'$)
while the production from two fermions via a $Z'$ is  $p$-wave suppressed.} We get
\begin{eqnarray}
 \sigma_{\rm ann} v_{\rm rel} &\approx & \nonumber
\frac{2 g_{B-L}^4}{{\pi M_{\rm DM}^2}} {\rm Re}\,\sqrt{1-\frac{M_{Z'}^2}{M_{\rm DM}^2}}+\\
&&+\, \frac{\lambda^2_{H\phi} v^2 \Gamma_h^*/M_{\rm DM}}{(4M^2_{\rm DM} - M_h^2)^2 + M_h^2  \Gamma_h^2}+\frac{\lambda^2_{H\phi}}{64\pi M^2_{\rm DM}} {\rm Re}\,\sqrt{1-\frac{M_h^2}{M^2_{\rm DM}}}\\
&&+\, \frac{\lambda^2_{S\phi} w^2 \Gamma_s^*/M_{\rm DM}}{(4M^2_{\rm DM} - M_s^2)^2 + M_s^2  \Gamma_s^2} 
 +\frac{\lambda_{S\phi}^2}{64\pi M_{\rm DM}^2}  {\rm Re}\,\sqrt{1-\frac{M_s^2}{M^2_{\rm DM}}}\nonumber.
\end{eqnarray}
where $\Gamma_{h,s}^*$ are the decay width into SM particles of a virtual $h,s$ with mass $2M_{\rm DM}$.\footnote{Longitudinal components of $W^\pm, Z$ enhance
$\Gamma_h^* \simeq 3 M_{\rm DM}^3/4\pi v^2$ at $M_{\rm DM}\gg M_h$,
such that 
$\sigma(\phi\phi^* \to h^* \to W^+W^-,ZZ) \simeq 3\sigma(\phi\phi^*\to hh)$ as demanded by $\SU(2)_L$ invariance.}
We neglected the $s/h$ interference. 
These cross sections are similar to the ones in the {DM scalar singlet} model~\cite{hep-ph/0702143,hep-ph/0011335}.


We now have all the ingredients to compute the DM density.
In fig.\fig{BmL} we plot it in the ($M_{Z'}, g_{B-L}$) plane, similarly  
to fig.\fig{m_g} for the $\SU(2)_X$ model.
However the U(1)$_{B-L}$ has a few extra free parameters,
most importantly the DM mass.
In fig.\fig{BmL} we thereby consider a few different values of the DM mass,
and assume that the extra free parameters are in ranges which give neither enhancements nor {cancellations in the various equations above}.
An important difference with respect to the previous model is that constraints from direct detection (in orange) are weaker.
Baryogenesis through  leptogenesis needs $T_{\rm RH}\circa{>}T_{\rm sph}$: in the plotted parameter region this is satisfied
when DM has a sizeable sub-thermal contribution, in addition to the super-cool contribution.

\section{Summary}\label{concl}
We presented a new mechanism that can reproduce the observed cosmological DM abundance
when DM is a weak-scale particle.
The mechanism arises in models where the weak scale is dynamically generated. 
The Universe remains trapped in a false vacuum where all particles are massless 
and undergoes a phase of thermal inflation during which all particles get diluted.
This phase can be ended by the QCD phase transition or by vacuum decay to the true vacuum, where
particles are massive.
Light particles are regenerated in the subsequent reheating phase, but the
DM abundance can remain suppressed, with a quite low temperature, due to supercooling.
Fig.\fig{evo} exemplifies the possible cosmological evolution.
When super-cooling ends at $T \sim \Lambda_{\rm QCD}$, the desired DM abundance is obtained for
weak-scale DM.

In section~\ref{specific} we have shown that super-cool DM is produced in a simple model proposed in~\cite{Hambye:2013sna},
where dynamical generation of the weak scale and DM stability is obtained adding to the SM a new $\SU(2)_X$ gauge group and
a new scalar doublet.
In section~\ref{B-L} we studied a model where the new gauge group is U(1)$_{B-L}$. 
In both models DM is reproduced dominantly through super-cooling for
DM masses of about 500 GeV and for DM couplings of order $10^{-4}$ ---
smaller than in the freeze-out scenario, such that the simplest  models of super-cool DM
are still allowed by direct detection.

The U(1)$_{B-L}$
gauge structure (and the scalar 
that breaks it), in addition of dynamically 
generating the symmetry breaking and of stabilizing DM, also 
gives rise to neutrino masses and to leptogenesis.
This is a welcome feature, given that super-cooling erases a possibly pre-existing baryon asymmetry,
which needs to be regenerated after reheating.
Depending on the model and on its parameter space, the reheating temperature can be either larger or smaller
than the decoupling temperature of weak sphalerons, such that the baryon asymmetry can be regenerated
either through leptogenesis or through cold baryogenesis, possibly during the phase transition that ends super-cooling.

\footnotesize

\subsubsection*{Acknowledgements}
This work was supported by the ERC grant NEO-NAT.
The work of T.H. and D.T. is supported by the Belgian FNRS-F.R.S., the ``Probing DM with neutrinos'' ULB-ARC grant, the ``be.h'' EOS grant n. 30820817, the IISN and a postdoctoral ULB fellowship.
We thank Pasquale Serpico for clarifications about~\cite{Iso:2017uuu} and Christian Gross for discussions and suggestions about cold baryogenesis.

\end{document}